\documentclass[conference]{IEEEtran}
\IEEEoverridecommandlockouts
\usepackage{cite}
\usepackage{amsmath,amssymb,amsfonts}
\usepackage{algorithmic}
\usepackage{graphicx}
\usepackage{textcomp}
\usepackage{graphicx}
\usepackage{xcolor}
\usepackage{verbatim}
\usepackage{multirow}
\usepackage{array}
\def\BibTeX{{\rm B\kern-.05em{\sc i\kern-.025em b}\kern-.08em
    T\kern-.1667em\lower.7ex\hbox{E}\kern-.125emX}}
\begin{document}

\title{Spark-MPI: Approaching the Fifth Paradigm of Cognitive Applications \\
}

\author{\IEEEauthorblockN{Nikolay Malitsky}
\IEEEauthorblockA{\textit{Brookhaven National Laboratory}}
\and
\IEEEauthorblockN{Ralph Castain}
\IEEEauthorblockA{\textit{Intel Corporation}}
\and
\IEEEauthorblockN{Matt Cowan}
\IEEEauthorblockA{
\textit{Brookhaven National Laboratory}}
}

\maketitle

\begin{abstract}
Over the past decade, the fourth paradigm of data-intensive science rapidly became a major driving concept of multiple application domains encompassing and generating large-scale devices such as light sources and cutting edge telescopes. The success of data-intensive projects subsequently triggered the next generation of machine learning approaches. These new artificial intelligent systems clearly represent a paradigm shift from data processing pipelines towards the fifth paradigm of composite cognitive applications requiring the integration of Big Data processing platforms and HPC technologies. The paper addresses the existing impedance mismatch between data-intensive and compute-intensive ecosystems by presenting the Spark-MPI approach based on the MPI Exascale Process Management Interface (PMIx). The approach is demonstrated within the context of hybrid MPI/GPU ptychographic image reconstruction pipelines and distributed deep learning applications.
\end{abstract}

\section{Introduction}
The fourth paradigm of data-intensive science, coined
by Jim Gray \cite{4P}, rapidly became a major conceptual approach
for multiple application domains encompassing and generating
large-scale scientific drivers such as fusion reactors and
light source facilities \cite{DOE1} \cite{DOE2}. Taking
its root from data management technologies, the paradigm emphasized
and generalized a data-driven knowledge discovery direction that
complemented the computational branch of scientific disciplines.
The success of data-intensive projects subsequently triggered
an explosion of numerous machine learning approaches
\cite{LeCun} \cite{Mnih}\cite{Hassabis} addressing
a wide range of industrial and scientific applications, such as
computer vision, self-driving cars, and brain modelling, just
to name a few. The next generation of artificial intelligent
systems clearly represents a paradigm shift from data processing
pipelines towards knowledge-centric applications. As shown in Fig. 1,
these systems broke the boundaries of computational and data-intensive
paradigms and began to form a new ecosystem by merging and extending
existing technologies. Identifying this trend as the fifth paradigm
aims to infer common aspects among diverse cognitive computing
applications and steer the development of complementary solutions
for addressing emerging and future challenges.

The initial landscape of data-intensive technologies was designed
after Google's Big Data stack over 15 years ago. It represented
a consolidated scalable platform bringing together database and
computational technologies. The open-source version of this platform
was further advanced with the Spark framework resolving the immediate
requirements of numerous data-intensive projects. The new model
of the Spark computing platform significantly extended the scope
of data-intensive applications, spreading from SQL queries to machine
learning to graph processing. According to
the data-information-knowledge-wisdom model \cite{Ackoff},
these projects eventually elevated data-information pipelines
to practical applications of knowledge development.
Cognitive systems of the fifth paradigm take the relay baton
from data-driven processing pipelines and generalizes
their scope with knowledge acquisition processes carried out by
rational agents through the exploration of their environments.

\begin{figure}[h]
	\centering
	\includegraphics[height=5.5cm]{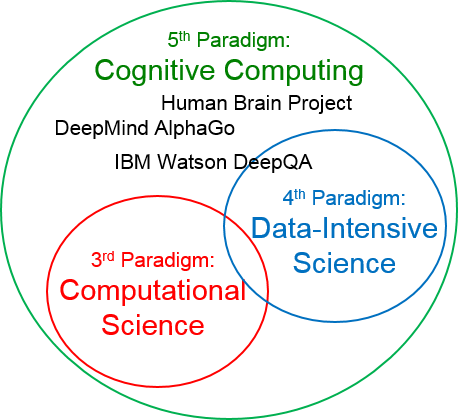}
	\caption{The Fifth Paradigm. The diagram shows the conceptual
		structure of the new paradigm integrating resources from
		both its ancestors, computational and data-intensive sciences,
		for building cognitive computing applications.}
\end{figure}

In contrast with the MapReduce embarrassingly parallel pipelines, machine
learning applications rely on the communication among distributed workers
for synchronizing their internal representations. This mismatch led
to the development of new distributed processing frameworks, such
as GraphLab \cite{GraphLab}, CNTK \cite{CNTK}, TensorFlow \cite{TensorFlow},
and Gorila \cite{Gorila}. As with any standard evolutionary spiral,
a variety and growing number of different approaches
eventually raised the question of their consolidation. Similar problems
are faced by researchers of large-scale scientific experimental facilities
and computational projects. Prior to the Big Data era, most scientific
algorithms were built within the third computational paradigm based
on HPC clusters and Message Passing Interface (MPI) communication model.
To address the immediate requirements of emerging applications,
the fourth data-intensive paradigm was developed by minimally
intersecting with the HPC ecosystem as shown in Fig. 1.

The strategic transition from data-intensive science towards
the fifth paradigm of composite cognitive computing applications
is a long-term journey with many unknowns. This paper addresses
the existing mismatch between Big Data and HPC applications
by presenting the Spark-MPI integrated platform aiming to bring
together Big Data analytics, HPC scientific algorithms and deep
learning approaches for tackling new frontiers of data-driven
discovery applications. The remainder of the paper is organized
as follows. Section 2 provides a brief overview of the Spark
data-intensive and MPI high-performance platforms and outlines
the Spark-MPI integrated approach based on the MPI Process
Management Interface (PMI). Section 3 and Section 4 further elaborate
the approach within the context of the hybrid MPI/GPU ptychographic
image reconstruction pipelines and distributed deep learning
applications. Section 5 provides insights into future directions
using the PMI-Exascale library. Finally, Section 6 and Section 7
survey related work and conclude with a summary.

\section{Spark-MPI Platform}

\begin{figure*}
	\centering
	\includegraphics[width=16cm,height=8cm]{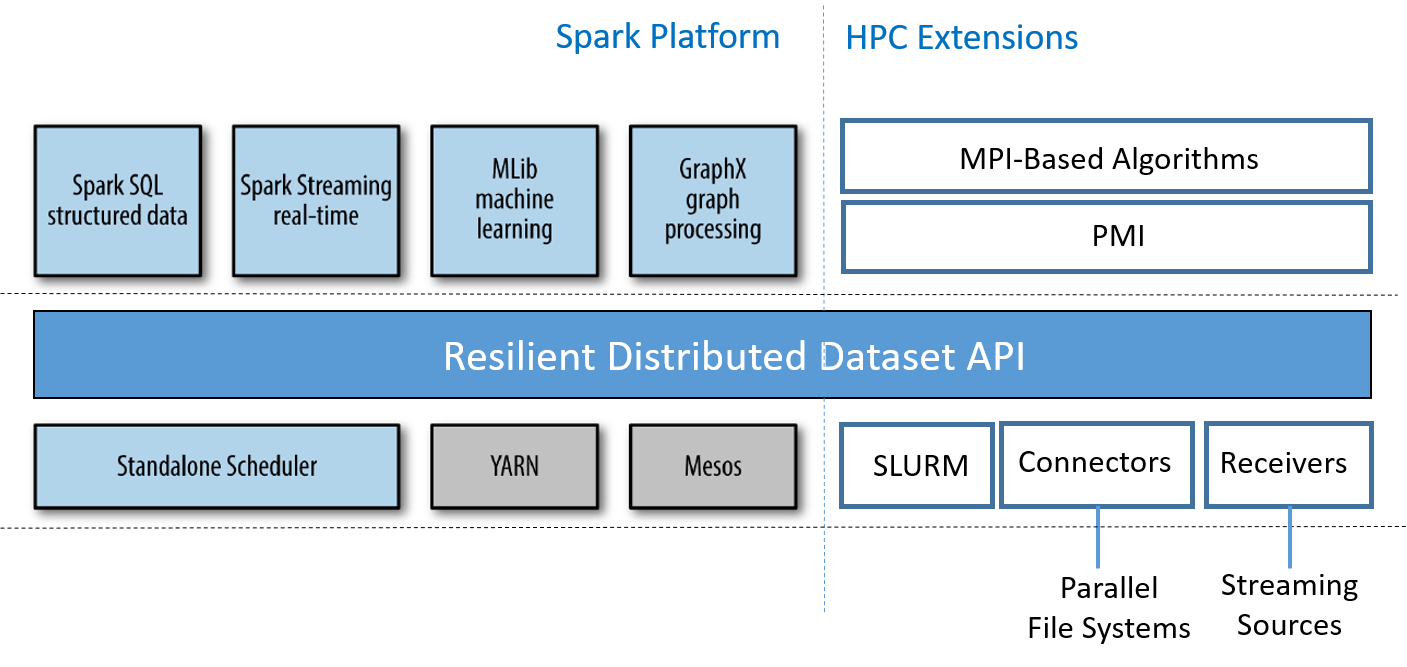}
	\caption{Conceptual architecture overview of the Spark-MPI platform}
\end{figure*}

The integration of data-intensive and compute-intensive ecosystems
was addressed by several projects. For example, Geoffrey Fox and
colleagues  provided a comprehensive overview of the Big Data
and HPC domains. Their application analysis \cite{Fox} was based
on several surveys, such as the NIST Big Data Public Working Group and
NRC reports, including multiple application areas: energy, astronomy
and physics, climate, and others. Overall, from a conceptual perspective,
the Spark-MPI platform can be considered as the Spark-based version
of the Exascale Fast Forward (EFF) I/O Stack three-tier
architecture \cite{EFF}. The Spark-MPI platform furthermore focuses
on the development of a common integrated approach addressing
a wide spectrum of applications including large-scale computational
studies, data-information-knowledge discovery pipelines for experimental
facilities, and reinforcement learning systems.

Fig. 2 shows a general overview of the Spark-MPI integrated environment.
It is based on the Spark Resilient Distributed Dataset (RDD)
middleware \cite{RDD} that decouples various data sources from
high-level processing algorithms. RDDs are distributed fault-tolerant
collections of in-memory objects that can be processed in parallel using
a rich set of operations, transformations and actions. The top layer
is represented by an open collection of high-level components addressing
different types of data models and processing algorithms, including machine
learning and graph processing. The interfaces between RDDs and distributed
data sources are provided by Connectors that are already implemented for
major databases and file systems.

In addition, Spark is designed to cover a wide range of workloads that
previously required separate distributed systems encompassing batch
applications, iterative algorithms, interactive queries, and streaming.
This combination forms a powerful processing ecosystem for building data
analysis pipelines and supporting multiple higher-level components
specialized for various workloads. The Spark Streaming
module \cite{SparkStreaming} further
extends the RDD pluggable mechanism with the Receiver framework enabling
to ingest real-time data into RDDs from streaming sources, such as Kafka
and ZeroMQ. Adherence to the RDD model automatically provided the Spark streaming
applications with the same functional interface and strong fault-tolerance
guarantees including exactly-once semantics.

The combination of a data-intensive processing framework with a consolidated
collection of diverse data analysis algorithms offered by Spark represents
a strong asset for its application in large-scale scientific projects across
different phases of the data-information-knowledge discovery path. In contrast
with existing data management and analytics systems, the Spark in-situ approach
does not require the transformation of data into different formats and provides
a generic interface between heterogeneous algorithms with heterogeneous data
sources. The current version of the Spark programming model, however,
is limited by the embarrassingly parallel paradigm and the Spark-MPI approach
serves to extend the Spark ecosystem with the MPI-based high-performance
computational applications.

MPI is an abbreviation for the Message Passing Interface standard that is
developed and maintained by the MPI Forum \cite{MPI}. The process of creating
the MPI standard began in April 1992 and as of now, it is used in most
HPC applications. The popularity of the MPI standard was determined by
the optimal combination of concepts and methods challenged by two
conflicting requirements: scope of parallel applications and portability
across different underlying communication protocols.

The MPI standard interface extends the Spark embarrassingly parallel
model with a rich collection of communication methods encompassing
Remote Memory Access (RMA), pairwise point-to-point operations
(e.g., send and receive), master-worker (e.g., scatter and gather)
and peer-to-peer (e.g., allreduce) collective methods. In addition,
the Barrier method within the collective category provides
a synchronization mechanism for supporting the Bulk Synchronous
Parallel paradigm. To address the scalability and performance aspects,
MPI introduced the concept of Communicators that defined the scope
for communication operations. As a result, this approach significantly
facilitated the development and integration of parallel libraries
using inter- and intra-communicators.

To support the MPI parallel model across different operating and
hardware systems, the MPI frameworks are based on a portable access
layer. One of its initial specifications, Abstract Device
Interface (ADI \cite{ADI}), was developed within the MPICH project.
Later, the MVAPICH project further extended the ADI implementations
to support InfiniBand interconnects and GPUDirect
RDMA \cite{GPUDirectRDMA}. The OpenMPI team introduced a different
solution,  Modular Component Architecture (MCA) \cite{MCA}, that was
derived as a generalization of four projects \cite{OMPI} bringing
together over 40 frameworks. MCA utilizes components (a.k.a. plugins)
to provide alternative implementations of key functional blocks such
as message transport, mapping algorithms, and collective operations.
As a result, OpenMPI Byte Transfer Layer (BTL) represents an open
collection of network-specific components for supporting shared memory,
TCP/IP, OpenFrabric verbs, and CUDA IPC, just to name a few.

Running parallel programs on HPC clusters requires interactions
with external process and resource managers, such as SLURM and Torque, to
enable the MPI processes to discover each other's communication endpoints.
Within the MPI ecosystem, this topic is typically addressed by the Process
Manager Interface (PMI \cite{PMI}).
While the implementation of the PMI specification
was never standardized, libraries nearly always consist of two parts: client
and server. The client code is linked with the MPI program and provides
messaging support to the server - it has no a priori knowledge of the
overall application, and must rely on the server to provide any required
information.

The PMI server is instantiated on each node that supports
an MPI process and has both the ability to communicate with
its peers (usually over an out-of-band Ethernet connection) and full
knowledge of
how to contact those peers (e.g., the socket upon which each peer is
listening). The server is typically either embedded in the local daemon of
the system's resource manager, or executed as a standalone daemon started
by a corresponding launcher such as $mpiexec$.

The Spark-MPI approach extends the scope of
the PMI mechanism for integrating the Spark
and MPI frameworks. Specifically, it complemented the Spark conventional
driver-worker model with the PMI server-worker interface for establishing
MPI inter-worker communications as outlined in Fig. 3 and Fig. 4.

\begin{figure}[h]
	\centering
	\includegraphics[height=3.8cm]{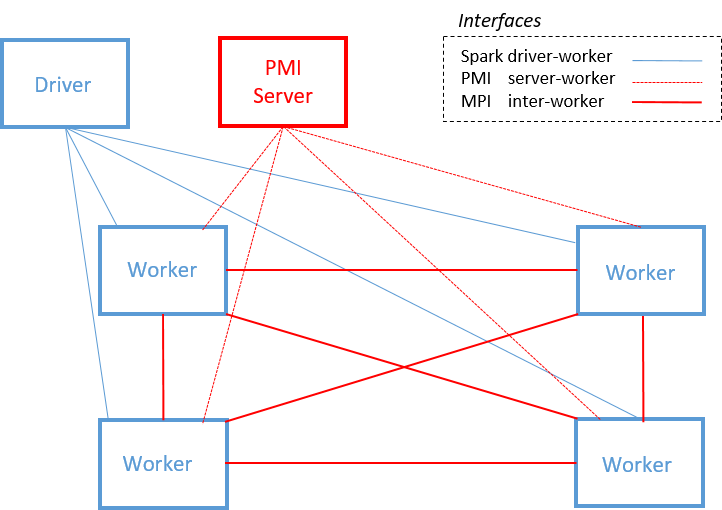}
	\caption{Spark-MPI approach}
\end{figure}

\begin{figure}[h]
	\centering
	\includegraphics[height=3.8cm]{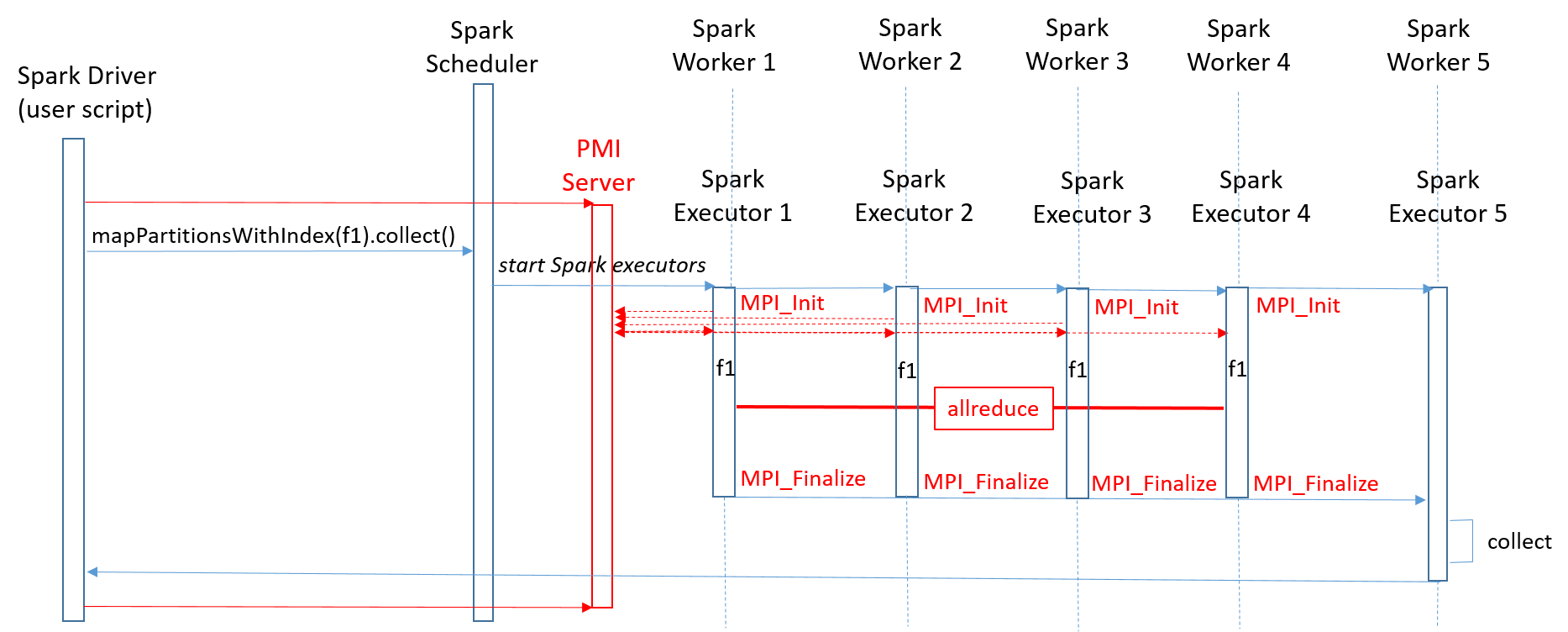}
	\caption{Sequence diagram of the Spark-MPI approach}
\end{figure}

The first version of the Spark-MPI approach was validated
with the primary internal process manager, Hydra,
used by two MPI projects (MPICH and MVAPICH).
The Hydra Process Manager (PM) is started
by the MPI launcher $mpirun$ on the launch node which subsequently
spawns a tree-based collection of intereconnected proxies on the
allocated nodes. Each proxy locally spawns one or more
application processes and then acts as the PMI server for
those processes. During initialization, each process ``publishes"
its connection information to the proxy, which then performs
a global collective operation to share the information across
all proxies for eventual distribution to the application.
Within the Spark-MPI integrated platform, MPI
application processes are started by the Spark scheduler (see Fig. 4).
Hydra local proxies therefore were modified to suppress their
launching functionality.

Recently, the Spark-MPI approach was integrated with
the Open MPI framework. The OpenMPI Modular Component Architecture
further streamlined its implementation as the $sparkmpi$ plugin of
the OpenRTE Daemon's Local Launch Subsystem (ODLS). The following
sections will demonstrate this approach within the context of
the hybrid MPI/GPU ptychographic image reconstruction pipelines and
deep learning applications.

\section{Ptychographic Image Reconstruction Pipelines}

Ptychography is one of the essential image reconstruction
techniques used in light source facilities. It was originally
proposed for electron microscopy \cite{Microscopy} and lately
applied to X-ray imaging \cite{Xray1} \cite{Xray2}. The method
consists of measuring multiple diffraction patterns by scanning
a finite illumination (also called the probe) on an extended
specimen (the object). The redundant information encoded in
overlapping illuminated regions is then used for reconstructing
the sample transmission function. Specifically, under the Born
and paraxial approximations, the measured diffraction pattern
for the jth scan position can be expressed as:
\begin{equation} \label{eq1}
\mathrm{I}_{j}(\mathbf{q}) = \mid \mathbf{F}\psi_{j} \mid^{2}
\end{equation}
where $\mathbf{F}$ denotes Fourier transformation, $\mathbf{q}$
is a reciprocal space coordinate, and $\psi_{j}$ represents
the wave at the exit of the object O illuminated by the probe P:
\begin{equation} \label{eq2}
\psi_{j} = \mathrm{P}(\mathbf{r}-\mathbf{r}_{j})\mathrm{O}(\mathbf{r})
\end{equation}
Then, the object and probe functions can be computed from
the minimization of the distance $\|\Psi - \Psi^{0} \|^{2}$
as \cite{ProbeRetrieval}:
\begin{equation} \label{eq3}
\epsilon = \|\Psi - \Psi^{0} \|^{2} = \Sigma_{j}\Sigma_{r} \mid \psi_{j}(\mathbf{r}) - \mathrm{P}^{0}(\mathbf{r}-\mathbf{r}_{j})\mathrm{O}^{0}\mathbf{r}\mid^{2}
\end{equation}
\begin{equation} \label{eq4}
\frac{\partial \epsilon}{\partial \mathrm{P}^{0}} = 0: \mathrm{P}^{0}(\mathbf{r}) = \frac{\Sigma_{j}\psi_{j}(\mathbf{r}+\mathbf{r}_{j})\mathrm{O}^{*}(\mathbf{r}+\mathbf{r}_{j})}{\Sigma_{j} \mid \mathrm{O}(\mathbf{r}+\mathbf{r}_{j})\mid ^{2}}
\end{equation}
\begin{equation} \label{eq5}
\frac{\partial \epsilon}{\partial \mathrm{O}^{0}} = 0: \mathrm{O}^{0}(\mathbf{r}) = \frac{\Sigma_{j}\psi_{j}(\mathbf{r})\mathrm{P}^{*}(\mathbf{r}+\mathbf{r}_{j})}{\Sigma_{j} \mid \mathrm{P}(\mathbf{r}-\mathbf{r}_{j})\mid ^{2}}
\end{equation}
These minimization conditions need to be augmented with the modulus
constraint \eqref{eq1} and included in the iteration loop. For example,
the comprehensive overview of different iterative algorithms is
provided by Klaus Giewekemeyer \cite{DifferenceMap1}. At this time,
the difference map \cite{DifferenceMap2} is considered as one of
the most generic
and efficient approaches to address these types of imaging problems.
It finds a solution in the intersection of two constraint sets
using the difference of corresponding projection operators,
$\pi_{1}$ and $\pi_{2}$, composed with associated maps,
$\mathrm{f}_{1}$ and  $\mathrm{f}_{2}$:
\begin{align} \label{eq6}
\psi^{n+1}&=\psi^{n}+\beta \Delta(\psi^{n}) \nonumber \\
\Delta &= \pi_{1}\circ \mathrm{f}_{2} - \pi_{2}\circ \mathrm{f}_{1}  \\
\mathrm{f}_{i}(\psi) &= (1+\gamma_{i})\pi_{i}(\psi) - \gamma_{i}\psi \nonumber
\end{align}
where $\gamma_{1,2}$ are relaxation parameters. In the context of
ptychographic applications, these projection operators are associated
with the modulus \eqref{eq1} and overlap \eqref{eq2} constraints.
By selecting different values of relaxation parameters,
the difference map \eqref{eq6} can be specialized to different variants
of phase retrieval methods and hybrid projection-reflection (HPR)
algorithms. Further developing HPR, Russel Luke \cite{RAAR} introduced
the relaxed averaged alternating reflections (RAAR) approach:
\begin{equation} \label{eq7}
\psi^{n+1} = [2\beta \pi_{0} \pi_{a} + (1-2\beta)\pi_{a} + \beta(1-\pi_{0})]\psi^{n}
\end{equation}
The RAAR algorithm was implemented in the SHARP program [31]
at the Berkley Center for Advanced Mathematics for Research
Applications (CAMERA).

SHARP is a high-performance distributed ptychographic solver
using GPU kernels and the MPI protocol. Since most equations with
the exception of \eqref{eq4} and \eqref{eq5} are framewise
intrinsically independent, the ptychographic application is
naturally parallelized by dividing a set of data frames among
multiple GPUs. Then, for updating a probe and an object,
the partial summations of \eqref{eq4} and \eqref{eq5} are
combined across distributed nodes with the MPI Allreduce
method as shown in Fig. 5.

\begin{figure}[h]
	\centering
	\includegraphics[height=4.0cm]{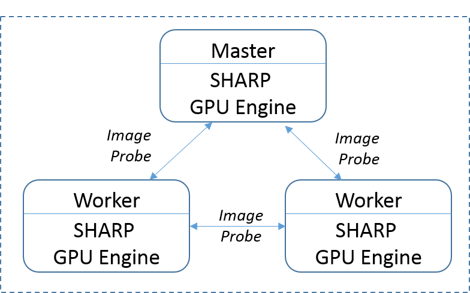}
	\caption{MPI communication model of the SHARP solver}
\end{figure}

The SHARP multi-GPU approach significantly boosted the performance
of ptychographic applications at the NSLS-II light source facility
and immediately highlighted the path for developing
near-real-time processing pipelines. Table I compares
the performance results for processing 512 frames on different
numbers of GPUs.

\begin{table}[htbp]
	\caption{Benchmark Results of the SHARP-NSLS2 Application }
	\begin{center}
		\begin{tabular}{|c|>{\centering}p{1.4cm}|>{\centering}p{1.4cm}|>{\centering}p{1.4cm}|}
			\hline
			\multirow{2}{*}{\textbf{Application}} & \multicolumn{3}{c|}{\textbf{Time (s) vs Number of GPUs (TESLA K80)}}
			\tabularnewline
			\cline{2-4}
			& \multicolumn{1}{c|}{1} & \multicolumn{1}{c|}{2} & \multicolumn{1}{c|}{4}
			\tabularnewline
			\hline
			SHARP-NSLS2 & 22.7 & 13.6 & 8.6
			\tabularnewline
			\hline
		\end{tabular}
	\end{center}
\end{table}

In the experimental settings,
the time interval between  frames takes approximately 50 ms, in other
words 25 seconds for 512 frames. And according to Table I,
the Spark-MPI application demonstrated the feasibility of
the near-real-time scenario. This direction is especially important
from the perspective of a new category of emerging four-dimensional
tomographic applications
that combine series of ptychographic projections generated at different
angles of object rotation. In these experiments, each ptychographic
projection is reconstructed from tens of thousands of detector frames
and the MPI multi-GPU version becomes critical for addressing
the GPU memory challenges.

The Spark-MPI integrated platform immediately provided a connection
between MPI applications and different types of distributed
data sources including major databases and file systems.
Furthermore, the Spark Streaming module reused and extended
the RDD-based batch processing framework with a new programming
abstraction called discretized stream, a sequence of RDDs,
processed by micro-batch jobs. These new batches are created
at regular time intervals. Similar to batch applications,
streams can be ingested from multiple data sources like Kafka,
Flume, Kinesis and TCP sockets.

For evaluating the Spark-MPI approach, the SHARP ptychographic
pipeline was tested with the Kafka streaming platform,
an Apache open source project that was originally developed
at LinkedIn \cite{Kafka}. The corresponding simulation-based 
scenario is described with a conceptual diagram (see Fig. 6).
\begin{figure}[h]
	\centering
	\includegraphics[height=4.0cm]{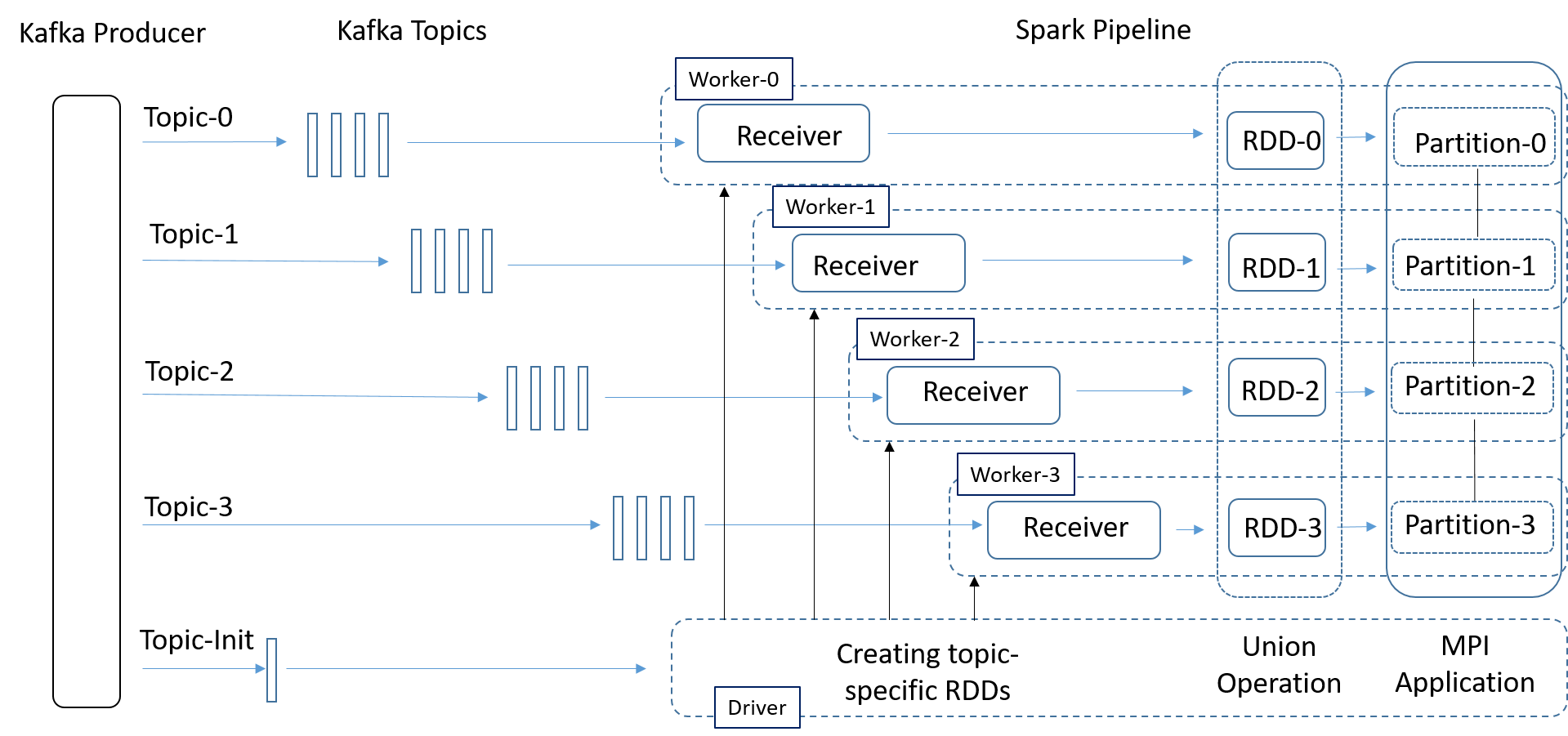}
	\caption{Streaming demo with the Spark-MPI approach}
\end{figure}

According to this scenario, the input stream represents
a sequence of micro-batches. The Spark driver waits for a topic-init
record and processes each micro-batch with the $run\_batch$ method.
At the beginning of this method, the Kafka data are ingested into
the Spark platform as the Kafka RDDs. To achieve a higher level of
parallelism, records of micro-batches are divided into topics that
are consumed by Kafka Receivers on distributed Spark workers.
Each Kafka Receiver creates a topic-specific RDD and the Spark driver
logically combines them together with a union operation. As a result,
it prepares a distributed RDD to be processed with the MPI application.

The acceleration of image processing algorithms with
the next generation of GPU devices further strengthened the direction
by creating the necessary conditions
for augmenting  photographic pipelines with optimization procedures.
The modern ptychographic approaches depend on many parameters and
their choice is important for achieving the most accurate reconstruction
results. For example, Fig. 7 and Fig. 8 demonstrate reconstructed
object phases for different choices of constraints.

\begin{figure}[h]
	\centering
	\includegraphics[height=4.0cm]{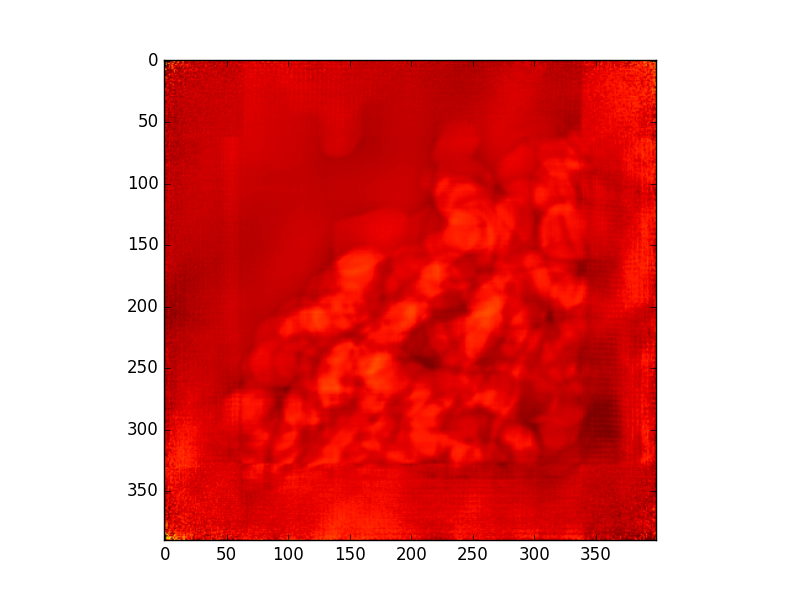}
	\caption{Object phases reconstructed from 40,000 frames (object
		constraints: amp\_max = 1.0, amp\_min = 0.0, phase\_max = $\pi/2$,
		phase\_min = - $\pi/2$)}
\end{figure}

\begin{figure}[h]
	\centering
	\includegraphics[height=4.0cm]{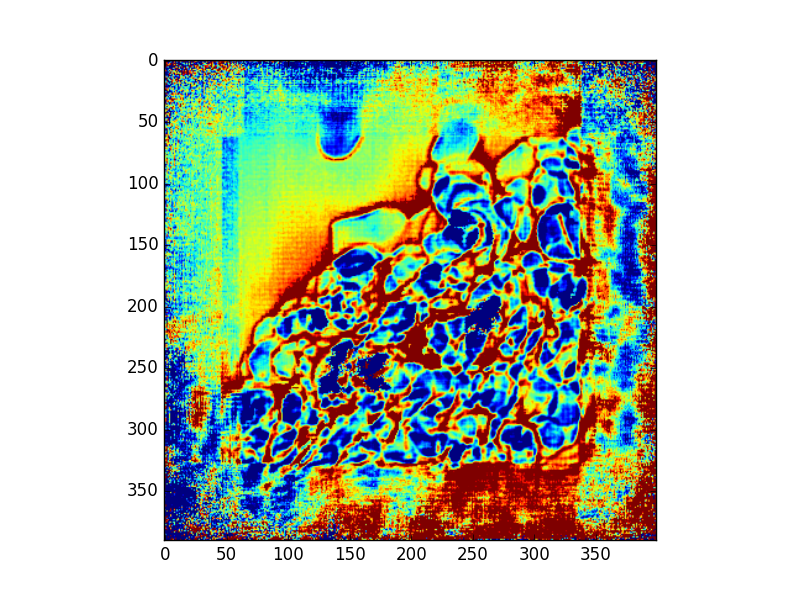}
	\caption{Object phases reconstructed from 40,000 frames (object
		constraints: amp\_max = 1.0, amp\_min = 0.95, phase\_max = 0.01,
		phase\_min = - 0.1)}
\end{figure}

Finding the
most optimal parameters can be automated with conventional
optimization approaches. In addition, the pipelines can be further
advanced with modern machine learning techniques for image analysis
and steering reconstruction algorithms.

\section{Deep Learning Applications}

The Spark-MPI platform was designed for building a new generation 
of composite data-information-knowledge discovery pipelines 
for experimental facilities. Deep learning applications advanced 
the scope and requirements of large-scale scientific projects 
to the next level. From the perspective of the fifth paradigm, 
Spark-MPI can be considered as a generic front-end of composite 
agent models for interacting with heterogeneous environments. 

Historically, distributed deep learning 
frameworks were developed within the fourth paradigm of 
data-intensive processing platforms. On the other hand, 
compute-intensive tasks have been already successfully 
addressed with a HPC stack of hardware and MPI applications 
from the third computational paradigm. The parallel acceleration 
of deep learning algorithms was then pursued by several 
MPI-based projects, such as CNTK \cite{CNTK}, 
TensorFlow-MaTex \cite{MaTex}, FireCaffe \cite{FireCaffe}, 
and S-Caffe \cite{SCaffe}. 

CNTK and TensorFlow are deep learning toolkits developed 
by Microsoft and Google, respectively. For distributed training, 
CNTK relies on the MPI communication platform and can be directly 
deployed on HPC clusters. In contrast, the original implementation 
of the TensorFlow distributed version is based on Google's gRPC 
interface developed for cloud computing systems using Ethernet. 
To leverage the HPC low latency interconnects, the TensorFlow-MaTEx 
project added two new TensorFlow operators, Global\_Broadcast and 
MPI\_Allreduce, and correspondingly modified the TensorFlow runtime. 
The FireCaffe and S-Caffe distributed approaches were developed around 
single-node Caffe deep learning solvers according to the data-parallel 
architecture. In addition, they further accelerated 
the data-parallel communication schema by replacing a parameter server 
with the allreduce communication pattern based on a reduction tree. 
Recently, the TensorFlow project was extended with a hybrid communication 
interface based on the combination of gRPC and MPI protocols. 
In contrast with other deep learning applications, the TensorFlow 
framework provides a pluggable mechanism for registering different 
communication interfaces that can be interchanged with other more 
advanced or application-specific versions.

Within the beamline composite pipeline platform, the Spark-MPI approach 
was  evaluated with Horovod \cite{Horovod}, a MPI training framework 
for TensorFlow. The Horovod team adopted Baidu's approach \cite{Baidu}
based on the ring-allreduce algorithm \cite{RingAllReduce} and 
further developed its implementation with the NVIDIA's NCCL library 
for collective communication. As a result, the ring-allreduce approach
replaced parameter servers of the TensorFlow distributed version 
with an efficient mechanism for averaging gradients among the TesnorFlow
workers. Their integration with the Horovod distributed framework 
consists of two primary steps as illustrated by the $horovod\_train$
method in Fig. 9. First, Horovod and MPI is initialized with $hvd.init$.
And then, the TensorFlow worker's optimizer is wrapped by $hvd.DistributedOptimizer$, a Horovod's ring-allreduce distributed 
adapter. 

\begin{figure}[h]
	\centering
	\includegraphics[height=4.5cm]{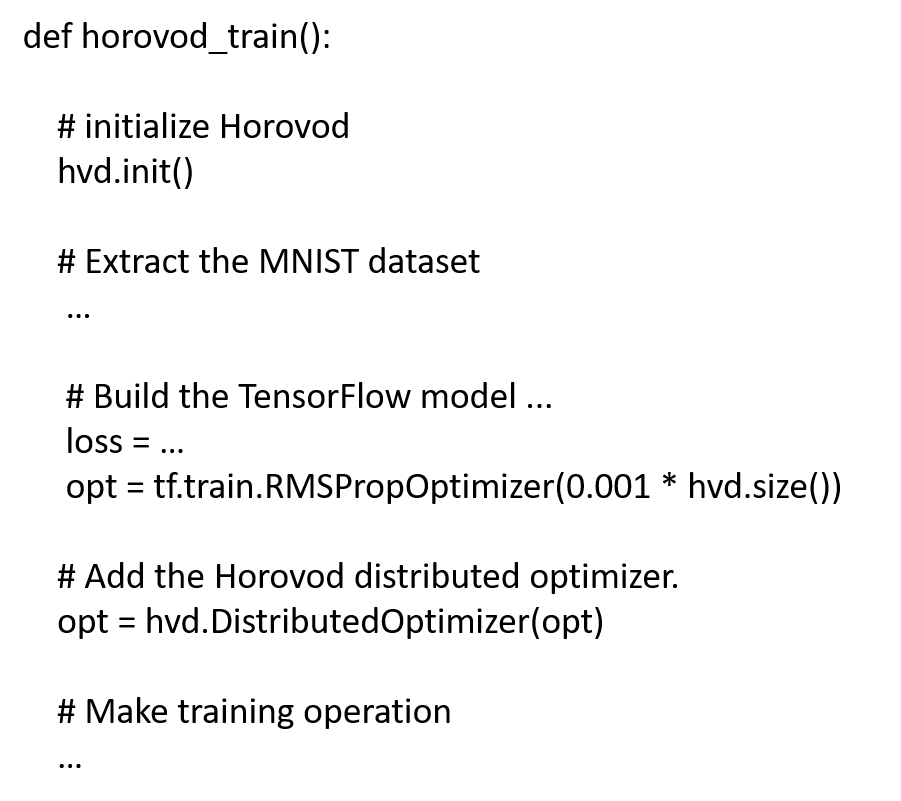}
	\caption{Horovod-TensorFlow example}
\end{figure}

The Spark-MPI pipelines enable to process the same method on Spark Workers 
within Map operations as shown in Fig. 10. To establish MPI communication among the Spark Workers, the Map operation needs 
only to define PMI-related environmental variables (such as 
PMIX\_RANK and a port number) 
for connecting the Horovod MPI application with the PMI server.

\begin{figure}[h]
	\centering
    \includegraphics[height=4.5cm]{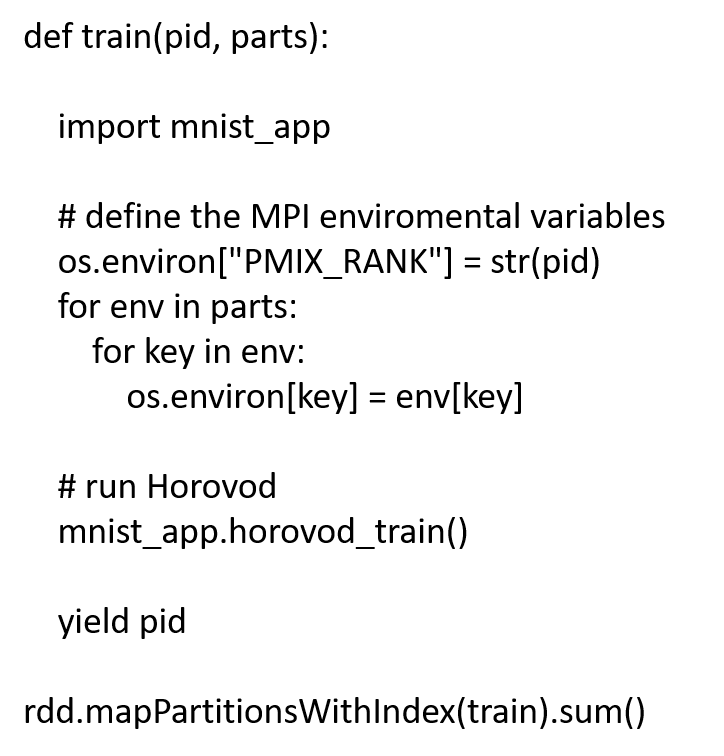}
    \caption{Spark-MPI-Horovod example}
\end{figure}

Implementing deep learning applications on the MPI parallel 
framework immediately extended the scope of the Spark-MPI ecosystem 
with composite pipelines as shown in Fig. 11. 

\begin{figure}[h]
	\centering
	\includegraphics[height=3.5cm]{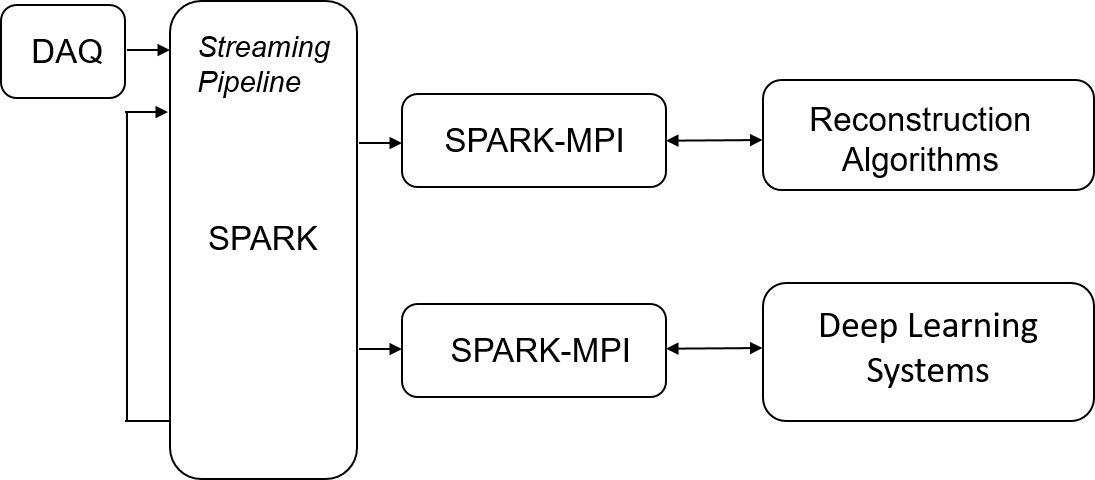}
	\caption{End-to-end machine learning pipeline including 
		reconstruction, image analysis, and feedback loop}
\end{figure}

For light source facilities, the development of composite pipelines 
involves two major topics: application of deep learning approaches 
for analyzing reconstructed images and development of machine 
learning feedback systems for steering reconstruction algorithms. 
According to the survey by 
Geert Litjens and colleagues \cite{MedicalImageAnalysisSurvey}, 
deep learning techniques pervade every aspect of medical image 
analysis: detection, segmentation, quantification, registration, 
and image enhancement. The feedback system can be viewed from 
the perspective of a rational agent that interacts with 
a reconstruction pipeline representing its environment. 
Depending on the applications, an agent 
can be built with different learning techniques. One of 
the most important 
breakthroughs is associated with the introduction of a deep 
Q-network (DQN) model for reinforcement learning \cite{Mnih}. 
The DQN-based 
approach demonstrated state-of-the-art results in various 
applications \cite{DRLSurvey} ranging from playing video 
games to robotics. 
As shown in Fig. 11, Spark-MPI  provides a generic front-end 
for distributed deep reinforcement learning platforms 
on the HPC cluster.

\section{Related Work}

The deployment of the Spark platform on HPC clusters and its comparison
with the MPI approaches has been addressed by several projects.
The Ohio State University team \cite{SparkRDMA}
proposed an RDMA-based design for the data shuffle of Spark over
InfiniBand. Alex Gittens and colleagues \cite{Gittens} demonstrated
the performance gap between a close-to-metal parallelized C version
and the Spark-based implementation of matrix factorization.
To resolve this gap, they introduced the Alchemist system for
socket-based interfacing between Spark and existing MPI libraries.
Michael Anderson and colleagues \cite{Anderson} proposed an alternative
approach based on the Linux shared memory file system.
The third solution suggested by Cyprien Noel, Jun Shi and Andy Feng
from the Yahoo Big ML
team extended the Spark embarrassingly parallel model with
the RDMA inter-worker communication interface. Later, this approach
was reused by the Sharp-Spark project \cite{SharpSpark} within
the context of ptychographic reconstruction applications. The Sharp-Spark
approach followed the Yahoo Big ML peer-to-peer model and augmented
it with a RDMA address exchange server that significantly facilitated
the initialization phase responsible for establishing Spark
inter-worker connections. As a result, the RDMA address exchange server
captured the PMI functionality of the MPI implementations and provided
a natural transition to the PMI-based Spark-MPI approach.

The similarity between the Spark driver-worker computational model and
the data-parallel approach of deep learning solvers triggered
the development of a new category of applications such as SparkNet \cite{SparkNet},
CaffeOnSpark \cite{CaffeOnSpark},
TensorFlowOnSpark \cite{TensorFlowOnSpark}, and BigDL \cite{BigDL}.
SparkNet directly relied on the Spark driver-executor scheme consisting
of a single driver and multiple executors running the Caffe or
TensorFlow deep learning solvers on its own subset of data.
In this approach, a driver communicates with executors for aggregating
gradients of model parameters and broadcasting averaged weights back
for subsequent iterations. According to the SparkNet-based benchmark,
the driver-executor scheme however introduced a substantial communication
overhead that was minimized by subdividing the optimization loop into
chunks of iterations. Addressing the same problem, the CaffeOnSpark team
proposed extending the Spark model with an inter-worker interface
providing a MPI Allreduce style method over Ethernet or InfiniBand.
Later, the same team began the TensorFlowOnSpark project based on 
their RDMA extension to the TensorFlow distributed platform.
In comparison with these projects, Spark-MPI aims to derive
an application-neutral mechanism based on the MPI Process Management
Interface for the effortless integration of Big Data and
HPC ecosystems.

\section{Path Towards Exascale Applications}

The validation of the Spark-MPI conceptual solution
established a basis for advancing this approach
towards the production programming model based on the
PMI-Exascale (PMIx) framework. Furthermore, this direction aligns
with proposed changes to the MPI standard \cite{MPIWG}
being supported by the PMIx community.

PMIx \cite{PMIx} was created in response to the ever-increasing
scale of supercomputing clusters, and the emergence of new programming
models such as Spark that rely on dynamically steered workflows. The
PMIx community has therefore focused on extending the earlier PMI work,
adding flexibility to existing APIs (e.g., to support asynchronous
operations) as well as new APIs that broaden the range of interactions
with the resident resource manager.

The initial version of the PMIx standard focused on resolving the scaling
challenges faced by bulk-synchronous programming models operating in
exascale systems\cite{PMIx2} \cite{PMIx3}. However, version 2 of
the standard directly addressed the needs of dynamic, asynchronous
programming models by providing APIs for changing resource
allocations (both adding and returning resources,
including the ability to ``lend'' resources back to the resource manager
for limited periods of time); controlling application execution (e.g.,
ordering termination and/or migration of processes, and coordinating
requests for application preemption); notification of events such as
connection requests and process failures; and connections to servers
from ``unknown'' processes not started by the server.

The Spark-MPI programming model utilizes the last feature as
a mechanism by which the processes started by the Spark scheduler
can connect to a local PMIx server. The PMIx library includes methods
for automatically authenticating connections to the server based on
a plugin architecture, thus allowing for ready addition of new methods
as required. Servers store their rendezvous information in files
located under system-defined locations for easy discovery, and the
client library executes a search algorithm to automatically find and
connect to a server during initialization.

Once connected to the server, the PMI-aware
processes can utilize PMIx to asynchronously request connections
to one or more processes. The connect and disconnect APIs
in version 2 of the PMIx Standard
retain support for bulk-synchronous programming models such as
today's MPI while providing the extensions needed for asynchronous
models. Both require that the operation be executed
as a collective, with all specified processes participating in the
operation prior to it being declared complete - i.e., all processes
specified in a call to {PMIx\_Connect} must call that API in order to
complete the operation. In addition, the
standard requires that the host resource manager (RM) treat the
specified processes as a new ``group'' when considering notifications,
termination, and other operations, and that no request to ``disconnect''
from a connected group be allowed to complete until all collectives
involving that group have also completed.

Finally, the PMIx community recognized that programming libraries
have continued to evolve towards more of an asynchronous model
where processes regularly aggregate into groups that subsequently dissolve
after completing some set of operations. These new approaches would
benefit from an ability to notify other processes of a desire to aggregate,
and to allow the aggregation process itself to take place asynchronously.

Accordingly, calls by PMI-aware processes to {PMIx\_Connect} 
are first
checked
by the PMIx server to see if the other specified participants are
available and have also called {PMIx\_Connect} - if so, then
the connection
request will result in
each involved process receiving full information about the
other participating processes (location, endpoint information, etc.)
plus a callback containing the namespace assigned to the connected
group. The latter can be considered the equivalent of a communicator
and used for constructing that object.

If one or more of the indicated processes has not executed its
call to {PMIx\_Connect}, then the server will issue an event notification
requesting that the process do so. Application processes can register
callback functions to be executed upon receipt of a corresponding event,
and events are cached so they can be delivered upon process startup
if the event is generated before that occurs. Once receiving a connection
request event, the process is given sufficient information in the
notification to allow it to join the requesting group, thereby
completing the collective operation. Applications can provide an optional
timeout attribute to the call to {PMIx\_Connect} so the operation will
terminate if all identified participants fail to respond within the
given time limit.

Note that any single process can be simultaneously engaged in multiple
connect operations. For scalability, PMIx does not use a collective to
assign a global identifier to the connect operation, instead utilizing
the provided array of process IDs as a default method to identify
a specific {PMIx\_Connect} operation. Applications can extend the ability
to execute multiple parallel operations by providing their own string
identifier for each collective as an attribute to the {PMIx\_Connect} API.
Note that all partitipants in a given collective are required
to call {PMIx\_Connect} with the same attribute value.

In cases where the involved hosts are controlled by different RMs,
the namespace identifier provided by the host RM for use in PMIx
is no longer guaranteed to be unique, thereby leading to potential
confusion in the process identifiers. Accordingly, PMIx defines
a method for resolving any potential namespace overlap by modifying
the namespace value for a given process identifier to include a
$cluster identifier$ - a string name for the cluster that is provided
by the host RM, or application itself in the case of non-managed
hosts.

The accumulated features of the PMIx distributed
framework are
identified as a new Exascale cluster service that supplements
the conventional resource management and scheduling platform
for gluing together HPC and Big Data applications.
On HPC clusters, support for PMIx
is currently integrated with the Open MPI
run-time environment and Simple Linux Utility for Resource
Management (SLURM \cite{SLURM}). Therefore, the deployment of the Spark-MPI
platform on HPC clusters will be streamlined
by adding SLURM into 
the list of Spark schedulers (see Fig. 2): Standalone, YARN \cite{YARN},
Apache Mesos \cite{Mesos} and Kubernetes \cite{Kubernetes}.
As illustrated by the Spark-MPI ptychographic and deep learning examples,
this deployment approach is consistent
with the Spark computational model. Furthermore, the asynchronous models
supported by the PMIx framework highlight the next direction for deploying
reinforcement learning architectures \cite{AsynDRL} on
HPC clusters.

\section{Conclusions}

The paper addresses
the existing mismatch between Big Data and HPC applications
by presenting the Spark-MPI integrated platform for bringing
together Big Data analytics, HPC scientific algorithms and deep
learning approaches for tackling new frontiers of data-driven
discovery applications. The approach was validated with 
three MPI projects (MPICH, MVAPICH and Open MPI) and established 
a basis for advancing the Spark-MPI interface towards 
the Exascale platform using the PMI-Exascale (PMIx) framework.
Furthermore, this direction aligns with a paradigm shift 
from data-intensive processing pipelines towards 
the fifth paradigm of 
knowledge-centric cognitive applications. Within the context of 
new applications, Spark-MPI aims to provide a generic 
front-end for distributed deep reinforcement learning platforms
on HPC clusters. As a result, the Spark-MPI 
platform represents a triple point solution located at 
the intersection of three paradigms.



\end{document}